\documentclass[aps,prb,twocolumn,groupedaddress,showpacs]{revtex4}
\usepackage{amsmath}
\usepackage{amssymb}
\usepackage{amsfonts}
\usepackage{graphicx}
\newcommand{\dprime}{\prime\prime}
\begin{document}

\title{Influence of Spin Wave Excitations\\ on the Ferromagnetic Phase Diagram in the
Hubbard-Model}

\author{W.~Rumsch}
\email[]{rumsch@physik.hu-berlin.de}
\author{W.~Nolting}
\affiliation{ Lehrstuhl Festk\"orpertheorie\\ Institut f\"ur Physik\\ Invalidenstrasse 110\\D-10115 Berlin\\Humboldt-Universit\"at zu Berlin\\
    Mathematisch-Naturwissenschaftliche Fakult\"at I}
\date{\today}

\begin{abstract}
The subject of the present paper is the theoretical description of collective electronic excitations, i.e. spin waves, in the Hubbard-model.
Starting with the widely used Random-Phase-Approximation, which combines Hartree-Fock theory with the summation of the two-particle ladder, we extend the theory to a more sophisticated single particle approximation, namely the Spectral-Density-Ansatz. Doing so we have to introduce a `screened` Coulomb-interaction rather than the bare Hubbard-interaction in order to obtain physically reasonable spinwave dispersions.
The discussion following the technical procedure shows that comparison of standard RPA with our new approximation reduces the occurrence of a ferromagnetic phase further with respect to the phase-diagrams delivered by the single particle theories.
\end{abstract}

\maketitle

\section{Introduction}\label{sec:sec1}
Recent publications \cite{Buen00,Okabe97,Singh98,ArKa99,KaAr00,Schu98,Costa01,BueGe01} on the theoretical description of spin waves in the Hubbard-model demonstrate an increased interest in the subject. The Hubbard-model, originally introduced in 1963 by Hubbard and others \cite{Hub163,Gutz163,Kanam63}, is one of the best investigated models in many-body-theory. It is commonly accepted to be the simplest model to hold for the itinerant as well as the localized character of the valence electrons in partially filled 3d-shells of transition and rare earth metals such as  Ni, Fe, Co. Despite the continuing interest only a handful of exact statements can be given on the possibility of a ferromagnetic groundstate.

Actually the model appears not to be very convenient for ferromagnetism. The exact solution of the $1-D$ Hubbard-model \cite{LiWu68} suggests the ground-state to be an antiferromagnetic one as well as the fact that in all dimensions the half-filled Hubbard-model can be mapped onto the antiferromagnetic Heisenberg-model in the strong coupling regime $\frac{U}{W}> 1$, where $W$ denotes the bandwidth and $U$ the interaction strength.
Furthermore the Mermin-Wagner-Theorem \cite{MerWa66} forbids the existence of magnetic long range order for $D<3$ at finite temperature $T$.
The zero-bandwidth limit excludes a ferromagnetic solution because $n_\uparrow = n_{\downarrow}$ is always valid \cite{Noll7}, herein $n_{\sigma}$ is the occupation number of $\uparrow$($\downarrow$)-spin.

If at all, we expect a ferromagnetic phase to occur in the strong coupling regime $\frac{U}{W}> 1$, away from exact half-filling in $D=3$.

The simple Hartree-Fock Approximation (HF) is able to deliver a ferromagnetic ground state in the Hubbard-model as long as the interaction strength is sufficiently large or respectively the density of states at the Fermi-surface $\rho_0({\epsilon_{\textrm F}})$ is considerably high, manifested in the famous Stoner-criterion $1 \leq U\rho_{0}(\epsilon_{\textrm F})$.

It is a well known fact that HF-theory overestimates the existence of spontaneous ferromagnetism. In particular the Curie-temperatures are orders of magnitude too high but nevertheless HF points in the right direction.

A more sophisticated approach beside others is the Spectral-Density-Ansatz (SDA) \cite{Noll72} based on Harris's and Lange's \cite{HarLa67} statements who have shown that the spectral density is dominated by two strongly developed peaks at the centre of the band $T_{0}$ and at $T_{0}+U$ when a finite hopping probability is taken into account. SDA basically is a two-pole ansatz of the form $S_{{\bf k}\sigma}(E-\mu)=\hbar\sum_{i=1,2}\alpha_{i\sigma}({\bf k}))\delta
  (E-E_{i\sigma}({\bf k}))$, where the in the first place unknown parameters, spectral weights $\alpha_{i\sigma}$ and peak positions $E_{i\sigma}({\bf k})$, are fitted via the exactly calculable spectral moments \cite{Noll72}.
The SDA-approximation results in a much more reliable
phase-diagram and more trustworthy Curie-temperatures than
HF-approximation. In a single particle theory, however, collective
excitations are not present. The study of these excitations in
itinerant ferromagnets requires the examination of the transversal
magnetic susceptibility. Collective excitations have a strong
influence on the stability of the ferromagnetic phase in
particular at higher temperatures since the excitations of magnons
tend to destroy the magnetic ordering. This is the reason for our
interest in describing collective excitations in the
Hubbard-model.

Facing the complexity of the many-body problem one usally has to
apply one or more suitable and reasonable approximations in order
to find physically interpretable results. Certain sequences of
such approximations will be presented in the following sections.
In Sect.~\ref{sec:sec2} we will develop the theory for the
magnetic susceptibility which results in the
Random-Phase-Approximation (RPA). RPA combines
ladder-approximation of the two-particle Green function with the
HF-approximation of the one-particle Green function and is
discussed in more detail in Sect.~\ref{sec:sec3}.

In Sect.~\ref{sec:sec4} we will combine ladder-approximation of
the two-particle Green function with the more sophisticated
SDA-approach of the one particle Green function which makes a
further approximation necessary, namely a screened interaction
$V_{eff}$, but will result in physically reasonable behaviour of
the collective mode. This is a new approach to the at the
beginning mentioned spin wave problem and discussed in detail. We
present our results in Sect.~\ref{sec:sec5} and discuss them with
respect to results of different authors and will finally close
with a short summary and an outlook in Sect.~\ref{sec:sec6}.
\section{Theory}\label{sec:sec2}
\subsection{Transversal Magnetic Susceptibility}\label{sec:sec2.1}
We use the so-called one band Hubbard-Hamiltonian where orbital degrees of freedom are neglected for simplicity. It reads in second quantized form in Fourier-space as
\begin{eqnarray}
  \label{eq:Hubbard_Hamilton_K}
  \hat{H}&=&\sum_{{\bf k}\sigma}(\epsilon({\bf k})-\mu)
a^+_{{\bf k}\sigma}a_{{\bf k}\sigma} +\nonumber\\
 &&\frac{1}{2}\frac{U}{N}\sum_{{\bf kpq}\sigma}a^+_{{\bf k}\sigma}
a_{{\bf k}+{\bf q}\sigma}a^+_{{\bf p}-\sigma}a_{{\bf p}-{\bf q}-\sigma}.
\end{eqnarray}
The operators $a^+_{{\bf k}\sigma}$ ($a_{{\bf k}\sigma}$) in (\ref{eq:Hubbard_Hamilton_K}) are the creation (annihilation) operators for fermions in ${\bf k}$-space.

Spin waves are collective excitations of the solid's state electronic system. The group of conduction electrons take on a coherent quantum state which is described as a gapless excitation with energy $E=\hbar \omega ({\bf{q}})$, where $\bf {q}$ is the wavevector of a so-called spinwave. The spin wave dispersion usually behaves as $E_{\bf q}\simeq D{\bf q}^2$ for small ${\bf q}$-values, where $D$ is the stiffness constant.

\begin{figure}[h]
    \includegraphics[width=\columnwidth]{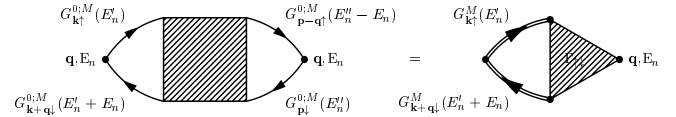}
    \caption{\label{fig:renorm}Representation of the polarisation propagator with help of the vertex function}
\end{figure}

In order to describe the excited system we make use of the retarded polarisation propagator and will analyze it with help of a diagramatic technique.
In our case the polarisation propagator is the transversal magnetic susceptibility. It is a two-particle Green function and given in energy representation by
\begin{equation}
  \label{eq:ret_chi}
  \chi^{+-}_{ij}(E)=-\frac{1}{\hbar^2}\langle\langle
  S^+_i;S^-_j\rangle\rangle_E
\end{equation}
where the spin-operators are
\begin{eqnarray}
  \label{eq:Suszop}
  S_{i}^+&=&\hbar a^+_{i\uparrow}a_{i\downarrow}\\
  S_{j}^-&=&\hbar a^+_{j\downarrow}a_{j\uparrow}.
\end{eqnarray}
Within Matsubara formalism \cite{Noll7,Matsub55}, which is a generalisation of the Green function formalism to the imaginary time scale,  this two-particle quantity is given by
\begin{eqnarray}\label{eq:ru:G2}
-\chi^{+-}_{q}(E_n)&=&\frac{1}{N}\sum_{\bf{kp}}\int\limits_{0}^{\hbar\beta}d\tau\exp({\frac{i}{\hbar}E_n\tau})\times\nonumber\\&&
\langle T_\tau[a_{\bf{k}\uparrow}^{+}(\tau)a_{\bf{k}+\bf{q}\downarrow}(\tau)a_{\bf{p}\downarrow}^{+}(0)a_{\bf{p}-\bf{q}\uparrow}(0)]\rangle\nonumber\\
&=&\hat{\chi}^{+-}_{q}(E_n)
\end{eqnarray}
The time-ordering operator $T_\tau$ in (\ref{eq:ru:G2}), for wich a generalized Wick's theorem exists \cite{Noll7}, sorts operators in a product of operators by their time-arguments. Matsubara-frequencies, denoted by subscript $n$, are given by
\begin{eqnarray}
  \label{eq:matsub_frequ}
E_n&=&\left\{ \begin{array}{r@{\quad:\quad}l}2n\,\frac{\pi}{\beta} & \mbox{bosons}\\(2n+1)\,\frac{\pi}{\beta} & \mbox{fermions}\end{array}\right..
\end{eqnarray}
One obtains the retarded form of (\ref{eq:ru:G2}), the actual
interesting quantity, simply by analytical continuation. The poles
of (\ref{eq:ru:G2}) yield the spin wave excitation spectrum.
Diagramatically (\ref{eq:ru:G2}) is represented by the graph given
on the left side of Fig.~\ref{fig:renorm}. The representation with
help of the vertex part (shaded triangle on the right of
Fig.~\ref{fig:renorm}) is very handy because it allows an
approximate partial summation.
The vertex part plays a similar role as the selfenergy in a single particle theory. The doubly lined propagators on the right side of Fig.~\ref{fig:renorm} indicate that we have to use the full single particle Green function $G^{M}_{{\bf k}\sigma}(E_n)$ instead of the free Green function $G^{0;M}_{{\bf k}\sigma}(E_n)$ indicated by singly lined propagators as on the left side of Fig.~\ref{fig:renorm}.

\begin{figure}[h]
  \includegraphics[width=\columnwidth]{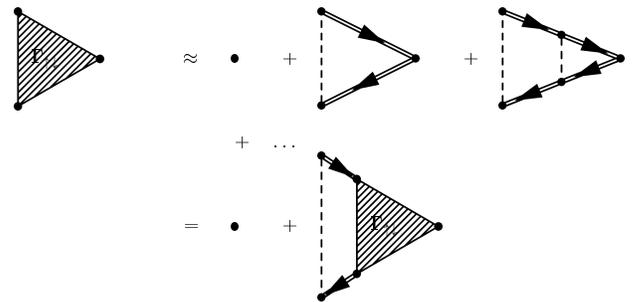}
  \caption{\label{fig:ladder}Partial summation of the ladder graphs}
\end{figure}

For the vertex part we have to find a practicable and physically reasonable approximation.
Summing up the ladder graphs (see Fig.~\ref{fig:ladder}) we find
with help of the diagram-'dictionary' \cite{Noll7}
\begin{eqnarray}
\Gamma_{\uparrow \downarrow}({\bf q},E_n;{\bf
  k},E^{\prime}_n)&=&1+\frac{-1}{\hbar^2\beta}\frac{U}{N}\sum_{\bf p}\sum_{E^{\dprime}_n}G^M_{{\bf p}-{\bf q}\uparrow}(E^{\dprime}_n-E_n)\nonumber\\
&&G^M_{{\bf p}\downarrow}(E^{\dprime}_n)\,\Gamma_{\uparrow\downarrow}({\bf q},E_n;{\bf p},E^{\dprime}_n).
\end{eqnarray}
The coupling parameter $U$ in the Hubbard-model is a constant which is why we can write
\begin{equation}
\Gamma_{\uparrow\downarrow}({\bf q},E_n;{\bf
  k},E^{\prime}_n)=\Gamma_{\uparrow\downarrow}({\bf q},E_n;{\bf p},E_n^{\dprime})\equiv \Gamma({\bf q},E_n)
\end{equation}
for the vertex part.

Introducing
\begin{equation}
\label{eq:lambda}
\Lambda_{\uparrow\downarrow}({\bf
  q},E_n)=-\frac{1}{{\hbar}^2 \beta}\frac{1}{N}\sum_{\bf k}\sum_{E_n^{\prime}}G^M_{{\bf k}\uparrow}(E_n^{\prime})G^M_{{\bf k}+{\bf q}\downarrow}(E_n+E_n^{\prime})
\end{equation}
we write for the whole propagator:
\begin{equation}\label{eq:n_vert}
\hat{\chi}^{+-}_{\bf q}(E_n)=\frac{\hbar\cdot \Lambda_{\uparrow\downarrow}({\bf q},E_n)}{1-U\cdot\Lambda_{\uparrow\downarrow}({\bf q},E_n)}.
\end{equation}
The abbreviation $\Lambda_{\uparrow\downarrow}({\bf q},E_n)$ on the right side of (\ref{eq:n_vert})
represents the polarisation propagator of zeroth order or pair bubble where no interaction line connects two vertex points.

Now, finding (\ref{eq:lambda}) solves the problem.
This task can be fullfilled by first rewriting (\ref{eq:lambda}) with help of the spectral representation of the Matsubara function, i.e. we use
\begin{eqnarray}
G^M_{{\bf k}\uparrow}(E^{\prime}_n)&=&\int\limits^{\infty}_{-\infty}dE_1\frac{S_{{\bf
      k}\uparrow}(E_1)}{iE^{\prime}_n-E_1}\nonumber\\
G^M_{{\bf k}+{\bf
    q}\downarrow}(E^{\prime}_n+E_n)&=&\int\limits^{\infty}_{-\infty}dE_2\frac{S_{{\bf k}+{\bf q} \downarrow}(E_2)}{i(E^{\prime}_n+E_n)-E_2}.
\end{eqnarray}
and write for $\Lambda_{\uparrow\downarrow}({\bf q},E_n)$:
\begin{eqnarray}
  \label{eq:prop4}
  \Lambda_{\uparrow\downarrow}({\bf
    q},E_n)&=&\frac{-1}{\hbar^2\beta}\frac{1}{N}\sum_{{\bf k}}\sum_{E^{\prime}_n}\int\limits^{\infty}_{-\infty}dE_1\int\limits^{\infty}_{-\infty}dE_2\times\nonumber\\&&\frac{S_{{\bf
      k}\uparrow}(E_1)}{iE^{\prime}_n-E_1}
\cdot\frac{S_{{\bf k}+{\bf q}
    \downarrow}(E_2)}{i(E^{\prime}_n+E_n)-E_2}.
\end{eqnarray}
For technical reasons we introduce two other abbreviations namely,
\begin{eqnarray}
  \label{eq:I}
F(E_n;E_1,E_2)=-\frac{1}{\hbar^2
  \beta}\sum_{E^{\prime}_n}\frac{1}{iE_n^{\prime}-E_{1}}&\times&\nonumber\\
\frac{1}{i(E^{\prime}_n+E_n)-E_{2}}
\end{eqnarray}
and
\begin{eqnarray}
  \label{eq:K}
K_{\bf q}(E_1,E_2)&=&\frac{1}{N}\sum_{\bf k}S_{{\bf k}\uparrow}(E_1)S_{{\bf k}+{\bf q} \downarrow}(E_2).
\end{eqnarray}
The polarisation propagator then becomes $\Lambda_{\uparrow\downarrow}({\bf q},E_n)=\iint\limits_{-\infty}^{\infty}dE_1dE_2F(E_n;E_1,E_2)\cdot K_{\bf q}(E_1,E_2)$.
Both expression (\ref{eq:I}) und (\ref{eq:K}) can be treated independently.

Evaluation of (\ref{eq:I}) requires the sum over Matsubara frequencies \cite{Noll7} which results in
\begin{equation}
  \label{eq:matsubsumme}
  F(E_n;E_1,E_2)=\frac{1}{\hbar^2}\left[ \frac{{\tilde f}_{-}(E_2)-{\tilde
      f}_{-}(E_1)}{iE_n-(E_2-E_1)}\right]
\end{equation}
where ${\tilde f}_{-}(E_{1,2})=\frac{1}{e^{\beta (E_{1,2})}+1}$.

The polarisation propagator $\Lambda_{\uparrow\downarrow}({\bf q},E_n)$ now looks with ${\tilde E}_{1,2}\longrightarrow (E_{1,2}-\mu)$ like
\begin{eqnarray}\label{eq:zwrgbns}
\Lambda_{\uparrow\downarrow}({\bf q},E_n)&=&\frac{1}{N\hbar^2}\sum_k\iint\limits^{+\infty}_{-\infty}dE_1dE_2\frac{\left[ f_{-}({\tilde E}_2)-f_{-}({\tilde E}_1)\right]}{iE_n-(E_2 - E_1)}\nonumber\\
&&\times S_{{\bf k}\uparrow}({\tilde E}_1)\cdot S_{{\bf k}+{\bf q} \downarrow}({\tilde E}_2).
\end{eqnarray}
Here $f_{-}({\tilde E}_{1,2})$ in the numerator is the Fermi function.

In order to represent the transversal susceptibility
(\ref{eq:ru:G2}) as a functional of a product of two single
particle spectral densities we performed a ladder approximation
which delivered the desired form (\ref{eq:zwrgbns}). Up to now we
did not make any statement about the one-particle spectral
densities in (\ref{eq:zwrgbns}). On account of the complexity of
the many-body problem we have to apply a suitable approximation
for the single particle spectral densities in (\ref{eq:zwrgbns}).
In the following section we will do exactly this with the
HF-approximation.
\section{Random-Phase-Approximation}\label{sec:sec3}
Evaluation of (\ref{eq:zwrgbns}) requires a three-dimensional
summation over wavevectors $\bf k$ and ${\bf k} + {\bf q}$
respectively.

The energy-dispersion for the conduction band in
Tight-Binding-Approximation is given for the $bcc$-lattice by
\begin{equation}
\epsilon({\bf k})^{bcc}=Z_1 T_1cos(k_x a)cos(k_y a)cos(k_z a)+T_0
\end{equation}
and for the $fcc$-lattice by
\begin{eqnarray}
\epsilon({\bf k})^{fcc}&=&4Z_1 T_1[cos(\frac{1}{2}k_x
a)cos(\frac{1}{2}k_y a)\nonumber\\ &&+cos(\frac{1}{2}k_y
a)cos(\frac{1}{2}k_z a)\nonumber\\ &&+cos(\frac{1}{2}k_z
a)cos(\frac{1}{2}k_x a)]+T_0.
\end{eqnarray}
The Hopping-integral $T_1$ for next nearest neighbours is given by
$T_1=-\frac{W}{2Z_1}$, where we used a bandwidth of $W=1eV$ for
all numerical calculations, $Z_1$ denotes the number of next
nearest neighbours ($Z_1^{bcc}=8$ and $Z_1^{fcc}=12$), the center
of the Band was choosen to be $T_0=0$. In order to reduce the
numerical effort to a minimum we use a similar $\bf q$-separation
procedure as given in \cite{Noll1} where the remaining $\bf
k$-summation can be transformed into a one-dimensional
energy-integration. This is due to the fact that the expression
\begin{equation}
\frac{1}{N}F(\epsilon({\bf k}))=\int F(x)\rho^{BDOS}(x)
\end{equation}

where
\begin{equation}
\rho^{BDOS}(x)\frac{1}{N}\sum_{\bf k}\delta(x-\epsilon({\bf k})
\end{equation}

is true for any $\bf k$-dependent function whose $\bf
k$-dependence is only given by a $\epsilon({\bf k})$-dependence.

\begin{figure}[h]
\centerline{
    \includegraphics[width=\columnwidth]{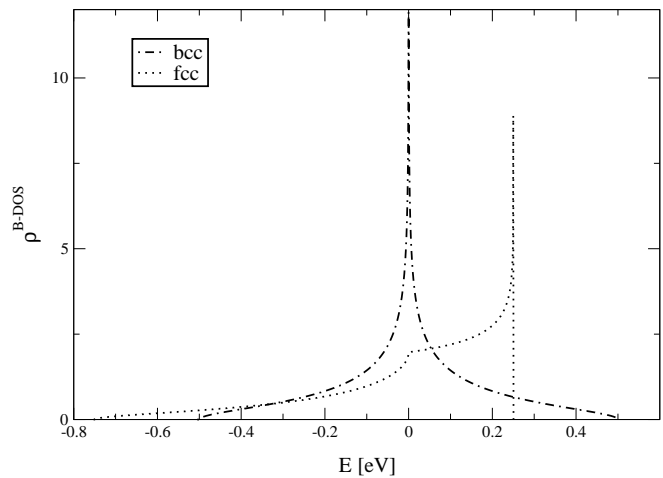}}
    \caption[]{\label{fig:BDOS}Bloch Density of States, bandwidth $W=1eV$, center of the band $T_0=0$}
\end{figure}

This procedure allows us to apply the numerical formulas given by
Jelitto \cite{Jel69} for the Bloch density of states (BDOS).
Picture \ref{fig:BDOS} shows the BDOS for the fcc and bcc-lattice
resulting from these formulas.

For the situation where ${\bf q}={\bf 0}$, however, the ${\bf
q}$-separation is not needed and one easily can check if there
exsists a gapless excitation with $\omega({\bf q})\longrightarrow
0$ as proposed by Lange \cite{Lange65}. Useful is the following
representation of the HF-spectral density
\begin{equation}
\label{eq:HF_specdens}
S_{{\bf k}\sigma}(E)=\hbar \delta(E-\epsilon ({\bf
  k})-U\langle n_{-\sigma}\rangle+\mu).
\end{equation}
The propagator (\ref{eq:zwrgbns}) reduces with (\ref{eq:HF_specdens}) and $E_{{\bf k},\sigma}=\epsilon_{\bf k}+U\langle n_{\sigma}\rangle$ to
\begin{equation}
\label{eq:lambda_no_TBA}
\Lambda_{\uparrow\downarrow}({\bf
  q},E_n)=\frac{1}{N}\sum_k\frac{f_{-}(E_{{\bf k}+{\bf q},\uparrow})-f_{-}(E_{{\bf k},\downarrow})}{iE_n-(E_{{\bf k}+{\bf q},\uparrow}-E_{{\bf k},\downarrow})}.
\end{equation}
For ${\bf q}={\bf 0}$ one easily can perform the summation over wave vectors because
\begin{eqnarray}
\langle n_{\sigma}\rangle&=&\frac{1}{N}\sum_{{\bf k}} f_{-}(E_{{\bf k},\sigma})=\frac{1}{N}\sum_{{\bf k}}\langle n_{\sigma}({\bf k})\rangle.
\end{eqnarray}
The condition for the denominator of (\ref{eq:n_vert}) to be zero is
\begin{equation}\label{eq:stoner_grenz}
1-U\cdot {\textrm Re}\Lambda_{\uparrow\downarrow}({\bf q}=0,E)=0
\end{equation}
or equivalently
\begin{eqnarray}
1-\frac{-U\cdot 2m}{E-U\cdot 2m +i0^+}&=&0
\end{eqnarray}
Here $m=\frac{1}{2}(n_{\uparrow}-n_{\downarrow})$ denotes the magnetisation.
One immediatly sees that this condition is fullfilled only for $E\equiv 0$ which means that using HF-theory a well
defined gapless excitation exsists.

Edwards \cite{Edwards62} already stated that excited states with one reversed spin in bandferromagnets must have an
energy of the form $E\sim \epsilon({\bf k}^{\prime})-\epsilon({\bf
  k})+J$ where $J$ is a ${\bf
  k}$-independent constant that contains the correlation energy. Clearly in HF-theory this is the case. In the event
  of transitions between majority- and minority-spin band the excitation energy is given by $\Delta E_{\uparrow\downarrow}({\bf k;{\bf q}})=\epsilon({\bf k}+{\bf q})-\epsilon ({\bf k}) + \Delta
E_{ex}$, where $\Delta E_{ex}=2mU$ is the rigid exchange splitting energy of the HF-theory.

\section{Modified Random Phase Approximation}\label{sec:sec4}
In this section we will use the SDA-theory as the single particle
input for the two particle propagator (\ref{eq:zwrgbns}). Compared
to HF-approximation SDA is a real many-body theory able to deliver
ferromagnetism in the Hubbard-model and tested by means of
comparison to different advanced many-body theories concerning
ferromagnetism in the Hubbard-model. Comparison of SDA,
Modified-Alloy-Analogy (MAA) and Modified-Perturbation-Theorie
with the numerical exact
Quantum-Monte-Carlo-Simulation\cite{Noletal} indicates that the
qualitative behaviour of the SDA phase-diagram is correctly
displayed. Furthermore are the Curie-temperatures orders of
magnitude lower than in HF-theory and therefore much more
reliable. HF-theory overestimates the occurence of ferromagnetism
in the Hubbard-model, although the tendency is right, in a
incredulous manner.

Though one cannot simply check the validity of the theory by
putting the SDA-selfenergy (or equivalently the spectral density)
for ${\bf q}={\bf 0}$ into the expression (\ref{eq:zwrgbns})
because the selfenergy in SDA consists of a Stoner or HF-like part
and a spin dependent rest. This spin-dependence prevents us from
 being able to check the validity in a similar manner as with
 HF-theory.

\begin{equation}
  \label{eq:SDA_Selbstenergie}
\Sigma_{{\bf k}\sigma}^{SDA}(E)=Un_{-\sigma}+\frac{U^2n_{-\sigma}(1-n_{-\sigma})}{E-B_{-\sigma}-U(1-n_{-\sigma})}.
\end{equation}
The spin dependence of the second term of
(\ref{eq:SDA_Selbstenergie}) is mainly given by the so called band
 shift $B_{-\sigma}$. This spin-dependence in SDA-theory is the main improvement and leads to a shift of
 $\uparrow$- and $\downarrow$-quasipartical-density of states
 (QDOS), for further details on the SDA see \cite{Noll1}.

However, the similarity of (\ref{eq:SDA_Selbstenergie}) to the
Stoner self energy let us hope that connecting SDA with ladder
approximation could work out and might even hold an improvement.
But numerical evaluation of (\ref{eq:zwrgbns}) using SDA-theory as
input in form of a sum of weighted $\delta$-functions shows
negative excitation energies which are unphysical.

With no apparent reason both approximations, the ladder
approximation for the two-particle propagator and the SDA for the
single particle propagator are not fully compatible.

The idea now was to introduce an effective Coulomb interaction
$V_{eff}$ in order to replace the Hubbard-U in (\ref{eq:n_vert})
and hopefully be able to shift the negative excitation energies to
the right energetic position. This was done numerically via the
condition
\begin{eqnarray}
V_{eff}&=&\left\{ \begin{array}{r@{\quad:\quad}l}U & if {\textrm
      Re}\Lambda_{\uparrow\downarrow}({\bf q}={\bf 0},E=0)=\frac{1}{U}
\\\ {\textrm Re}\frac{1}{\Lambda_{\uparrow\downarrow}({\bf 0},0)}\nonumber & else\end{array}\right.,
\end{eqnarray}
Indeed, replacing the Hubbard-$U$ in the polarisation propagator
(\ref{eq:n_vert}) with $V_{eff}$ given by the above condition the
unphysical behaviour could be repaired. The susceptibility
(\ref{eq:n_vert}) now is to look at as
\begin{equation}
\label{eq:MRPA}
  \hat{\chi}^{+- (MRPA)}_{\bf q}(E_n)=\frac{\hbar\cdot \Lambda^{(SDA)}_{\uparrow\downarrow}({\bf q},E_n,U)}{1-V_{eff}\cdot\Lambda^{(SDA)}_{\uparrow\downarrow}({\bf q},E_n,U)}.
\end{equation}
The term $\Lambda^{(SDA)}$ in (\ref{eq:MRPA}) is
(\ref{eq:zwrgbns}) with the appropriate input of the SDA-single
particle density of states.

The introduction of the effective or screened interaction of
course is to rate as yet another approximation. But the
interaction in a two particle theory not necessarily has to be the
same as the interaction in a single particle theory. It is the
essence of a two particle theory to take additional correlations
into account. HF-theory for instance neglects the correlation
between particles of opposite spin. This can be seen by the
diagramatical representation of the self-energy in HF-theory,
where the exchange-interaction part does not contribute to the
self-energy. The only contribution to the HF-self-energy comes
from the direct Coulomb-term, wich describes the interaction of an
electron with an effective field caused by the other particles in
the system. The contribution coming from the exchange-interaction,
however, is taken into account in ladder approximation as can be
seen from (\ref{eq:lambda}). Diagramatically spoken we did nothing
else than another renormalization in order to be able to use
SDA-theory. The broken lines in Fig.~\ref{fig:ladder} now has to
be imagined as broken double lines indicating an effective
interaction in the vertex.

\begin{figure}[h]
\centerline{
    \includegraphics[width=\columnwidth]{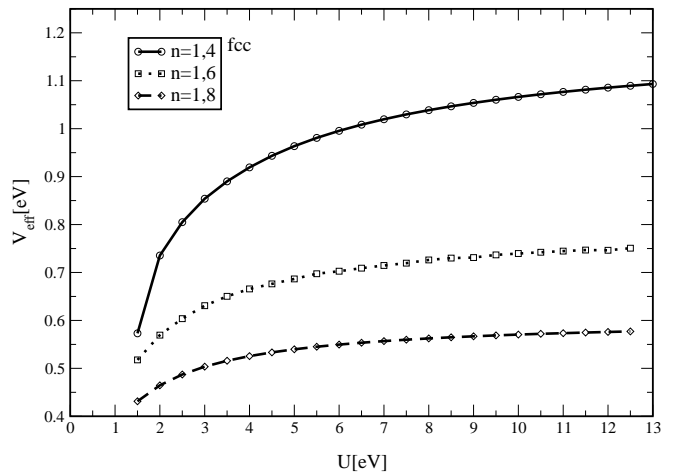}}
    \caption[]{\label{fig:VofUfcc}$V_{eff}=f(U)$ for band occupations $n=1.4$, $n=1.6$, $n=1.8$ at $T=0K$}
\end{figure}

The effective potential as a function of the bare Hubbard-$U$ can be seen in Fig.~\ref{fig:VofUfcc} for the $fcc$-lattice at $T=0K$ and three different band-occupations.
Since the effective potential seems to saturate for larger values of $U$ we restricted ourselves to $U$-values up to $7eV$.

\begin{figure}[h]
\centerline{
    \includegraphics[width=\columnwidth]{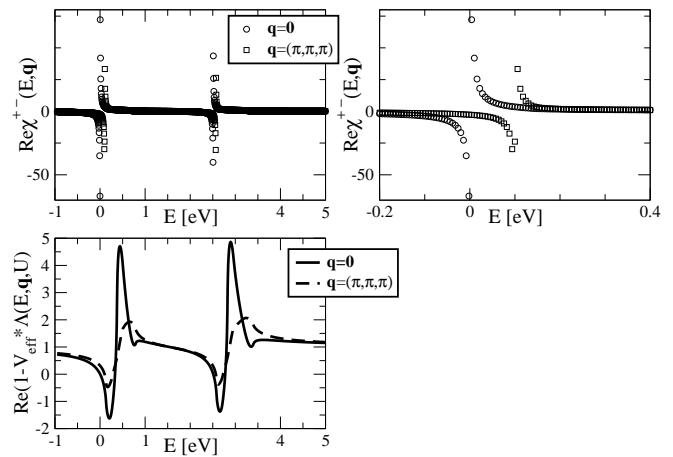}}
    \caption[]{\label{fig:SDA}Upper graphs:
      $Re\hat{\chi}^{+-(MRPA)}_{\bf q}(E_n)$ for ${\bf q}={\bf 0}$ and ${\bf q}=({\bf \pi,\pi,\pi})$,
      the two-peak structure of
      the two-particle spectral density is clearly visible in the upper left graph; the lower
      graph shows $Re(1-V_{eff}\Lambda_{\uparrow\downarrow}({\bf
        q},E_n))$ to indicate the poles of (\ref{eq:n_vert})}
\end{figure}

The SDA-QDOS is mainly given by two wheighted $\delta$-like peaks.
This two-peak structure is conserved in our MRPA-method. This can
be seen from Fig.\ref{fig:SDA}, where
$Re\hat{\chi}^{+-(MRPA)}_{\bf q}(E_n)$ and
$Re(1-V_{eff}\Lambda_{\uparrow\downarrow}({\bf q},E_n))$ are
plotted for ${\bf q}={\bf 0}$ and ${\bf q}={\bf\pi}$ as a function
of Energy.
The effect of the SDA is that the upper set of ${\bf q}$-dependent
curves, i.e. the one shifted by $\sim U$ as aresult of the
SDA-input, shows a dispersion too. This set could be identified as
optical modes which are not present in RPA-theory. Calculations
for the real substances Fe \cite{Cooke285} and Ni \cite{Cooke185},
however, indicate that a separation between acoustical and optical
modes about $\sim U$ is one order of magnitude to large. That is
why we only did look at the low lying excitation mode.
\section{Results and Discussion}\label{sec:sec5}
In this section we present the results of MRPA and RPA. The latter one we used as the established `test` theory for
our purposes.

In MRPA we restricted ourselfs to the $fcc$-lattice since an asymmetric quasi particle density of states (Q-DOS)
supports the tendency towards a ferromagnetic alignement of the spins in the Hubbard-model \cite{AOKA00}.

\begin{figure}[h]
\centerline{
    \includegraphics[scale=.35]{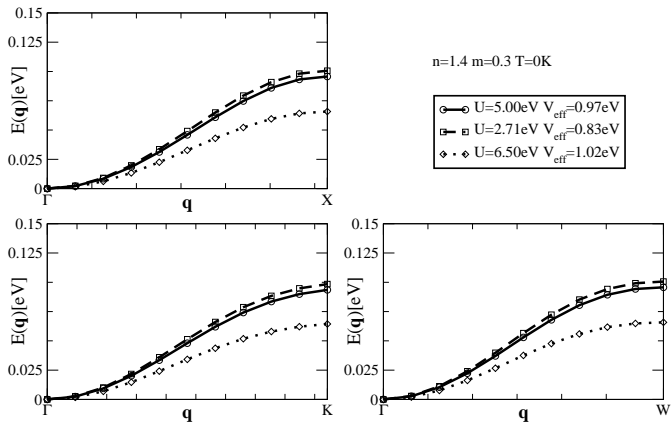}}
    \caption[]{\label{fig:SDA_EofqofU_symm}Magnon dispersion in MRPA along different high symmetry directions of the
    FBZ for the $fcc$-lattice, parameter: $U=2.71eV;5.00eV;6.50eV$ at fixed band occupation $n=1.4$, $T=0K$, see the
     bordered box for values of $V_{eff}$}
\end{figure}

Figure \ref{fig:SDA_EofqofU_symm} shows isotropic spin wave dispersions as a function of $U$ for the $fcc$-lattice
in MRPA. The dispersion curves were computed along different directions of high symmetry in the First-Brillouin-Zone
(FBZ). This isotropy for the $fcc$-lattice is in agreement with results observed using Gutzwiller-ansatz \cite{Buen00}.

\begin{figure}[h]
\centerline{
    \includegraphics[width=\columnwidth]{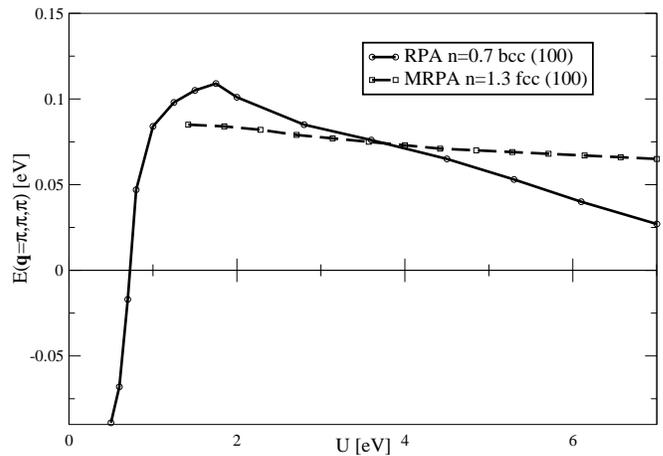}}
    \caption[]{\label{fig:SDA_EofX} Behaviour of magnon energies at the edge of the FBZ as a function of $U$ for RPA
    and MRPA at $T=0K$, qualitativ similar behaviour for larger $U$-values in both theories}
\end{figure}

Figure \ref{fig:SDA_EofqofU_symm} also shows the behaviour of the stiffness $D$ in dependency of the correlation
 parameter $U$. For increasing values of $U$ the magnon energy at the edge of the FBZ weakens or in other words $D$
 decreases with increasing $U$. In Fig.~\ref{fig:SDA_EofX} $E({\bf q}=(\pi, \pi, \pi))=f(U)$ is plotted for RPA and
 MRPA for fixed band occupations in either case. We see that in RPA theory the magnon energy at the edge of the FBZ
 continuously increases up to a maximum and afterwards decreases with higher values of $U$. In MRPA the magnon energy
 at the edge of the FBZ decreases continously with increasing $U$. This is in contrast to a statement given by B\"unemann
 et.al.\cite{BueGe01} where the authors found the opposite behaviour using (multiband) RPA namely increasing $D$ with
 increasing $U$. The behaviour of decreasing $D$ with increasing $U$ on the other hand was observed by B\"unemann et.al.
 \cite{Buen00} using their recently developed multiband Gutzwiller-ansatz.

We rather think that the decreasing $D$ with increasing $U$-behaviour is an intrinsic feature of the Hubbard-model.

\begin{figure}[h]
\centerline{
    \includegraphics[width=\columnwidth]{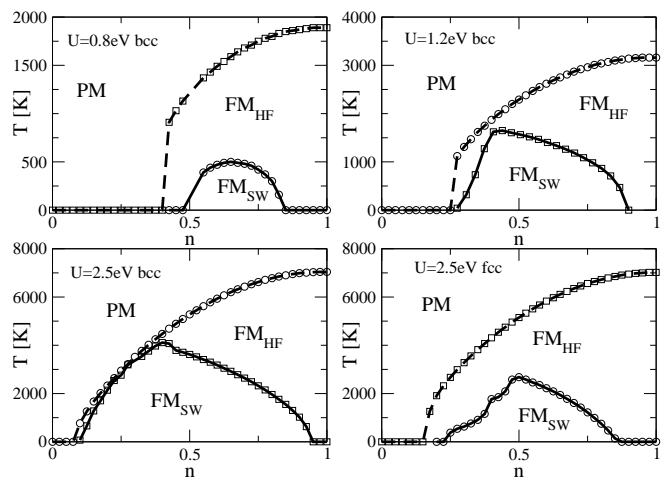}}
    \caption[]{\label{fig:HF_Tcofn+Tvonn}RPA: $T_{\textrm C}(n)$-phase diagrams,
      the upper curves are $T_C (n)$-phase diagrams computed with
      single-particle HF-theory denoted by subscript HF, in each case the lower curves indicate the phases where spin
      waves are excited denoted by subscript SW}
\end{figure}

To us the most surprising result is that using RPA we could compute a qualitatively rather new $T_{\rm C}(n)$-phase
diagram which is to see in Fig.~\ref{fig:HF_Tcofn+Tvonn}.

Figure \ref{fig:HF_Tcofn+Tvonn} compares HF-phase diagrams with the corresponding phases where the spin wave mode,
indicated by subscript SW, exsists for three interaction strengths $U$ for the $bcc$-lattice. One example shows a
similar behaviour for the $fcc$-lattice.

Numerically we distinguished the SW phase by looking at the stiffness or the magnon energy at the edge of the FBZ
respectively. This means that whenever the stiffness $D$ becomes negative the ferromagnetic state breaks down.

The dramatic influence of the spin wave mode on the stability of the ferromagnetic phase is obvious from
Fig.~\ref{fig:HF_Tcofn+Tvonn}. Since Stoner-theory overestimates the possibility of ferromagnetic ordering
in the Hubbard-model the lack of the single particle theory is compensated by the corresponding two-particle
 theory and even results in new Curie-temperatures.

This feature of the RPA is also described  for the case of Neel-temperatures \cite{Singh98}.

\begin{figure}[h]
\centerline{
    \includegraphics[width=\columnwidth]{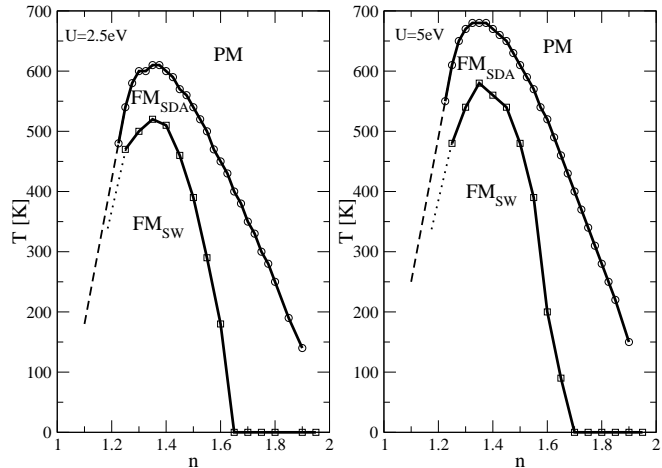}}
    \caption{\label{fig:SDA_Tcofn}MRPA: Comparison of single and two-particle phase diagram, in each case the
    upper curve is the $T_{\textrm C}(n)$-curve from SDA-theory denoted by subscript SDA, the lower curves are
    the `collective` phases denoted by subscript SW, fcc-lattice}
\end{figure}

In the modified theory MRPA the influence of the collective mode on the shape of the $T_{\textrm C}(n)$-phase
 diagram can be seen from Fig.~\ref{fig:SDA_Tcofn}.
The resulting phase diagram again is physically understandable.
The excitation of spin waves reduces Curie-temperatures and the region of band occupation but the qualitative
shape more or less stays the same as in the single particle theory.

The ferromagnetic phase for both theories in general is reduced
due to weakening of the spin wave dispersion close to half filling
and the temperatur driven increased magnon excitation. From figure
\ref{fig:SDA_Tcofn} can be seen that the $T_{\textrm C}$-curves
have their peaks for $n\sim 1.4$. This feature is due to the fact
that the bandoccupation that favours a ferromagnetic alignment of
the spins in the SDA-approach is mainly given by the
lattice-structure. For the fcc-lattice the favourable
bandoccupation is $n> 1$ and for $n\sim 1.4$ the ferromagnetic
groundstate is almost saturized. The concrete value for
$T_{\textrm C}$ is mainly given as a function of the correlation
strength $U$.

Comparison with $T_{\textrm C}(n)$-phase diagrams of different single particle approximations \cite{Noletal}
including SDA and the numerically exact Quantum-Monte-Carlo-Simulation \cite{Ul98} demonstrate that besides
 Curie-temperatures the region of band occupation where ferromagnetism exists is reduced. This indicates that
 our `collective`-phase diagrams are much more reliable then the ones computed using the corresponding single
 particle theories.
\section{Summary and Outlook}\label{sec:sec6}
In the present paper we computed spin wave dispersions and $T_{\textrm C}(n)$-phase diagrams using standard RPA
theory and a modified theory where a reliable approximation for a single particle Green function is used as an
input for the two particle Green function. Although we did not follow a stringent mathematical procedure for our
MRPA as described exemplary for the case of RPA in \cite{BaKa61} we were able to show that our results using the
modified theory MRPA gives physically reasonable results. These results are comparable to results which we get
 using standard RPA and to results of a different method namely the Gutzwiller-ansatz as described in \cite{Buen00}.
 Although RPA is a widely used standard technique \cite{Okabe97,Singh98,ArKa99,KaAr00,Schu98,Costa01} to our knowledge
 a `complete` $T_{\textrm C}(n)$-phase diagram was not published before.

Our aim was to get a deeper insight into the phenomenon of collective excitations in a model bandferromagnet.
To a `first` approximation our results show that just counting on a single particle theory in order to describe
the problem of collective magnetism is not informative enough. Collective excitations strongly influence the
stability of a ferromagnetic state and one might use `collective`-phase diagrams as tests for single particle theories.

If one wanted to improve the whole theory one had to find an approximation for the vertex function better, i.e.
 more informative, than ladder approximation and an approximation for the single particle selfenergy at the same
  theoretical level. Parquet-Ansatz in connection with Coherent-Potential-Approximation CPA as described for the
  electrical conductivity in the Anderson model \cite{Janis} might be a good starting point. A CPA-type
  approximation be able to hold for ferromagnetism in this respect is the Modified-Alloy-Analogy (MAA) \cite{Herrmann}
   which already yields an improved phase-diagram compared to the SDA-approach \cite{Noletal}.

\end{document}